\documentclass[runningheads]{llncs}
\usepackage{graphicx}
\usepackage{algorithmic}
\usepackage{cite}
\usepackage{amsmath,amssymb,amsfonts}
\usepackage{algorithmic}
\usepackage{tablefootnote}
\usepackage{caption}
\usepackage{subfig}
\usepackage{textcomp}
\usepackage{epstopdf,url,color,array}
\usepackage{multirow} 
\usepackage[ruled,vlined]{algorithm2e}

\begin{document}
	\title{Stain Style Transfer of Histopathology Images Via Structure-Preserved Generative Learning}
	%
	%
	\author{Hanwen Liang \inst{1} \and Konstantinos N. Plataniotis \inst{1} \and Xingyu Li \inst{2}}
	\institute{The Edward S. Rogers Department of Electrical and Computer Engineering, University of Toronto \and Electrical and Computer Engineering, University of Alberta}
	%
	%
	%
	\maketitle              
	\begin{abstract}
		Computational histopathology image diagnosis becomes increasingly popular and important, where images are segmented or classified for disease diagnosis by computers. While pathologists do not struggle with color variations in slides, computational solutions usually suffer from this critical issue. To address the issue of color variations in histopathology images, this study proposes two stain style transfer models, SSIM-GAN and DSCSI-GAN, based on the generative adversarial networks. By cooperating structural preservation metrics and feedback of an auxiliary diagnosis net in learning, medical-relevant information presented by image texture, structure, and chroma-contrast features is preserved in color-normalized images. Particularly, the smart treat of chromatic image content in our DSCSI-GAN model helps to achieve noticeable normalization improvement in image regions where stains mix due to histological substances co-localization. Extensive experimentation on public histopathology image sets indicates that our methods outperform prior arts in terms of generating more stain-consistent images, better preserving histological information in images, and obtaining significantly higher learning efficiency. Our python implementation is published on https://github.com/hanwen0529/DSCSI-GAN.
		
		\keywords{Stain style transfer\and generative model\and color normalization\and structural similarity\and computational histopathology.}
	\end{abstract}
	
	\section{Introduction}
	\label{introduction}
	Computational histopathology is a promising field where image processing and machine learning techniques are applied to histopathology images for disease diagnosis. One critical issue with it is color variation among histopathology images. Due to the variation in chemical stains and staining procedures, tissue slides can differ greatly from each other in visual appearance. Other factors may also introduce variation to visual appearance, including the storage conditions of stain prior to use and the handling of slides. Since color information is recognized as key factor in an automatic histopathology analysis system, the tissue slices’ varying appearance directly increases diagnosis complexity and impacts the quality and accuracy of a computational solution\cite{b5}.
	To address the color variation issue in computational histopathology, one potential, practical solution is to eliminate color variation in the pre-learning stage and many color normalization solutions are proposed. Briefly, prior color normalization approaches can be categorized into three groups: (i) histogram matching methods \cite{i8,g8,h0}; (ii) spectral matching based on stain decomposition \cite{b9,h0,b22,b23}; (iii) style transfer based on generative learning \cite{k90,c0,MICCAI2019,ISBI2019,MLMIR2019}. Among the three categories, histogram matching methods treat color distribution independent of image content, thus may introduce image distortions after normalization. The spectral matching methods are based on Beer-Lambert law for stain decomposition\cite{g21} and achieve good performance for light-absorbing stains. However, there are many scattering stains in histopathology images that do not follow Beer-Lambert Law. To overcome these limitations, deep learning methods, especially generative adversarial networks(GAN) \cite{p0}, are exploited, hoping for a generalizable color normalization solution for computational histopathology.
	
	In this study, we develop a novel stain-style transfer framework combining a GAN network and a classification network for color normalization on histopathology images. 
	The proposed framework learns the histopathological staining protocol from training set and achieves stain style transformation with high efficiency. To preserve histopathological information delivered by texture, structural, and color content in images, we respectively innovate the use of structural similarity index matrix (SSIM) \cite{e4} and directional statistics based color similarity index (DSCSI)\cite{lee2015towards} as the image reconstruction loss function in training process. A feature preservation loss function is exploited in the proposed framework to minimize the loss of discriminative representations in images. 
	We perform extensive experimentation to evaluate the proposed SSIM-GAN and DSCSI-GAN models against prior arts. The results suggest that the proposed models succeed in transferring stain style, generating stain-consistent images, and bringing significantly higher learning efficiency than prior arts.
	
	In summary, our contributions are in two-folds. First, we propose two stain style transfer models that learn histopathological staining protocols and realize color normalization. Second, we introduce the use of SSIM and DSCSI metrics in GAN's learning, preserving structural information in images when transferring color patterns. DSCSI-GAN is the first to utilize image chromatic spatial organization to regularize GAN in stain style transfer and achieves noticeable normalization improvement in image regions where chemical stains mix due to histological substances overlap.

\section{Method}
\label{methodsec}
\begin{figure*}
	\centering
	\includegraphics[width=\textwidth]{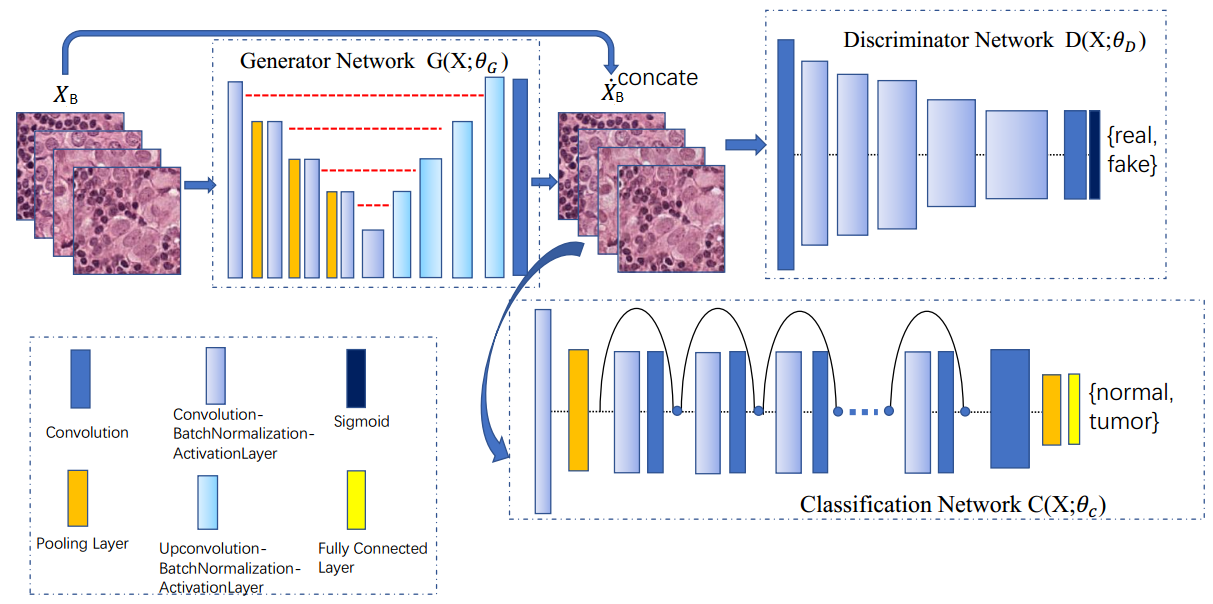}
	\caption{Overview of the stain-style transfer network and classification network. }
	\label{model}
\end{figure*}
	
	To clarify the histopathological stain style transfer problem, we define the dataset of histopathology images from pathology lab A as ${{X}_{A}}=\{x_{A}^{1},\ldots x_{A}^{n}\}$ and its corresponding labels ${{Y}_{A}}=\{y_{A}^{1},\ldots y_{A}^{n}\}$, where $x_{A}^{i}\in {{R}^{3}}$ is a color image in RGB format and $y_{A}^{i}\in (0,1)$ indicates whether it is normal or tumor image. We also define the test dataset from pathology lab B as ${{X}_{B}}$ which represents the same type of tissue but has different staining appearance. The proposed stain normalization model aims at generating images ${{\dot{X}_{B}}}\,$ from source images ${{{X}_{B}}}\,$, where ${{\dot{X}_{B}}}\,$ should preserve the histological information in ${{{X}_{B}}}\,$ and have the same color style as training set ${{{X}_{A}}}\,$. 
	
	\subsubsection{Network Architectures:}
	The block diagram of the proposed framework is shown in Fig.~\ref{model}. In this study, we deploy an auxiliary model $C(X;{{\theta }_{C}})$ to augment the learning procedure of GAN \cite{p0} for color normalization. A GAN model comprises of a generator $G(X;{{\theta }_{G}})$ and a discriminator $D(X;{{\theta }_{D}})$, which respectively generates color-normalized images and discriminates original images from generated ones. The model learns color characteristics within an entire dataset ${{{X}_{A}}}\,$ in training and finally is capable of processing query images ${{{X}_{B}}}\,$ so that the normalized images ${{\dot{X}_{B}}}\,$ appear the same stain style of ${{{X}_{A}}}\,$. We cooperate image structural similarity loss functions in GAN's learning for histological information preservation.

	As presented in Fig.~\ref{model}, the proposed framework is composed of three convolutional neural networks. 
	For generator $G(X;{{\theta }_{G}})$, we adapt U-Net \cite{m0} to communicate image histological content between layers and to embrace precise localization property. 
	For discriminator $D(X;{{\theta }_{D}})$, we adopt the discriminator part from vanilla GAN \cite{p0}, which takes real images or fake images generated by $G(X;{{\theta }_{G}})$ for classification. 
	Our framework also deploys ResNet \cite{n0} as the auxiliary net $C(X;{{\theta }_{C}})$ to classify histopathology images as tumor or normal images. After the network is trained on ${{X}_{A}}$ , we use $C(X;{{\theta }_{C}})$'s feature map before the last fully-connected layer as feedback for GAN's update. 

	\subsubsection{Objective Functions:}
	Given the proposed GAN based stain transfer framework with parameter set $\theta = \{{\theta }_{G},{\theta }_{D},{\theta }_{C}\}$, where ${{\theta }_{G}}$ and ${{\theta }_{D}}$ are the trainable parameters of generator and discriminator of GAN and ${{\theta }_{C}}$ is the trainable parameters of the diagnosis net, the objective function $L(\theta)$ of the proposed model is composed of three loss functions:
	\begin{equation}
	L(\theta) = \alpha {{L}_{GAN}}(G,D)+\beta {{L}_{reco}}(G) +\gamma {{L}_{fp}}(G,C)),
	\label{lossFunction}
	\end{equation}
	where $\alpha ,\beta ,\gamma$ are the weights to balance GAN loss ${{L}_{GAN}}(G,D)$, image reconstruction loss ${{L}_{reco}}(G)$, and feature preserving loss ${{L}_{fp}}(G,C)$. 
	
	The GAN loss ${{L}_{GAN}}(G,D)$ \cite{p0} drives the system to perform an adversarial game (i.e. the generator learns to generate images that can fool discriminator whose task is to distinguish between real and fake images). 
	Since the generic GAN loss ${{L}_{GAN}}(G,D)$ may induce the model to generate an image losing histological patterns, we introduce two more loss functions in training. 
	
	The reconstruction loss ${{L}_{reco}}(G)$ measures the difference between generated images ${{\dot{X}_{A}}}\,$ and its original counterpart ${{X}_{A}}$, and enforces the generator learning image color distribution and maintaining the structural information in images at the same time. Specifically, structural information refers to the knowledge about the structure of objects, e.g. spatially proximate, in the visual scene \cite{Wang2004}. Particularly in the context of computational histopathology, structural information mainly refers to the spatial organization of histological substances, i.e. multicellular structures, in histopathology images. Such information is a key for downstream computational histopathology and thus should be maintained in color normalization. In prior arts, generative networks usually adopt MSE as image reconstruction loss function. However, MSE-driven models are prone to generating a smooth/blur reconstruction where some structural information in the original signal is missing \cite{zhao2017loss}. To address this problem, we introduce two loss functions based on image structural similarity (i.e. SSIM and DSCSI) to measure quality of generated images. The motivations behind is that structural similarity correlates well with human’s perception of image quality\cite{o12} and facilitates the networks to maintain the texture and structural patterns in images.
	
	The SSIM based reconstruction loss function when training $G(X;{{\theta }_{G}})$ and $D(X;{{\theta }_{D}})$ can be formulated by
	\begin{equation}
	L_{\text{reco}}(G)=E_{x\sim {X_{A}}}[1-SSIM(x,G(x;\theta_{G}))],
	\label{SSIMLoss}
	\end{equation}
	where ${L}_{\text{reco}}(G) \in [0,1]$ and SSIM \cite{e4} is the structural similarity index matrix between original image ${x}_{A}^{i}$ and generated image $\dot{x}_{A}^{i}$ by $G(X;{{\theta }_{G}})$. 
	As SSIM is proposed for gray-scale images, in practice, we first map RGB images to gray-scale images. A sliding window is applied to obtained gray-scale images and image differences within the sliding windows, characterized by luminance, structure, and contrast, are evaluated and averaged for a single SSIM value. We name the stain style transfer model with SSIM as SSIM-GAN in this study.
	
	Note SSIM is proposed for grayscale image quality measurement and may fail for color images that exhibits chromatic deviations \cite{lee2015towards}. Aware that histopathology images are color in nature and color normalization in these images focuses on chromatic style transfer, we also develop a DSCSI based reconstruction loss and replace the SSIM metric in GAN learning. The correpsonding model is called DSCSI-GAN in this paper. The DSCSI loss combines chromatic and achromatic similarity and is formulated as:
	\begin{equation}
	L_{\text{reco}}(G)=E_{x\sim {X_{A}}}[1-DSCSI(x,G(x;\theta_{G}))].
	\label{DSCSILoss}
	\end{equation}
	In training, the original image and generated image are first transformed to the S-CIELAB space\cite{zhang1997spatial} and six similarity measures in the hue/chroma/lightness channels are developed to compute the similarity score in SDCSI loss.
	
	The last term in Eqn. (1) is used to preserve discriminative histological features in images. In training, we feed original image ${{{x}_{A}}}\,$ and generated image ${{\dot{x}_{A}}}\,$ to the pre-trained auxiliary diagnosis net $C(X;{{\theta }_{C}})$ and extract feature representations after the final convolution layers. Following Cho's work \cite{c0}, we obtain feature-preserving loss by means of Kullback-Leibler divergence.
	
	\subsubsection{Training Procedure:}
	Our proposed methods are composed of two learning stages. First, a diagnosis net used as the auxiliary classifier $C(X;{{\theta }_{C}})$ is trained on the training set ${{{X}_{A}}}\,$ and image labels. 
	Second, the GAN type model is trained following Algorithm~\ref{algorithm} to optimize the proposed objective functions. 
	\begin{algorithm*}
		\caption{Training GAN model with Minibatch Stochastic Gradient Decent.}
		\label{algorithm}
		\begin{algorithmic}
			\STATE {\bfseries Input:} data ${{{X}_{A}}}\,$,${{{Y}_{A}}}\,$, pre-trained model $C(X;{{\theta }_{C}})$ \\
			Initialized the weights of networks $G(X;{{\theta }_{G}})$,
			$D(X;{{\theta }_{D}})$
			
			\FOR{number of training iterations}
			\STATE Sample minibatch of 2m images, half normal images and the other half tumor images: 
			$x_{A}^{1},x_{A}^{2},x_{A}^{3},\ldots x_{A}^{2m}$;
			\STATE Feed images to $G(X;{{\theta }_{G}})$ and generate $\dot{x}_{A}^{1}, \dot{x}_{A}^{2}, \dot{x}_{A}^{3},\ldots \dot{x}_{A}^{2m}$;
			\STATE Update $D(X;{{\theta }_{D}})$ by ascending its stochastic gradient:
			
			$\nabla_{\theta_D}{1\over 2m}\sum_{i=1}^{2m}\left(\log D\left(x_A^i;\theta_D \right)+\log\left(1-D\left(\dot{x}_A^i;\theta_D\right)\right)\right)$;
			\STATE Feed original images $x_{A}^{1},x_{A}^{2},x_{A}^{3},
			\ldots x_{A}^{2m}$ and generated images $\dot{x}_{A}^{1}, \dot{x}_{A}^{2}, \dot{x}_{A}^{3}, \ldots \dot{x}_A^i$ to $C(X;{{\theta }_{C}})$ to obtain feature representations $F(x_{A}^{i})$ and $F(\dot{x}_A^i)$;   
			
			\STATE Update $G(X;{\theta }_{G})$ by descending its stochastic gradient: 
			
			$\nabla_{\theta_{G}}{1\over 2m}\sum_{i=1}^{2m}\left(\log\left( 1-D\left(\dot{x}_A^i;\theta_D \right)\right)+l_{\text{reco}}(x_A^i,\dot{x}_A^i)+\mbox{KL}\left[F\left(x_A^i\right)||F\left(\dot{x}_A^i\right)\right]\right)$
			\ENDFOR
			\STATE {\bfseries Output:}Networks model $G(X;{{\theta }_{G}})$,
			$D(X;{{\theta }_{D}})$
		\end{algorithmic}
	\end{algorithm*}
	
	\section{Experimentation}
	\label{experimentsec}
	\subsubsection{Dataset:}We conduct the experiment based on the Camelyon16 dataset \cite{dataset}, which is composed of 400 WSIs from two different institutes, Radbound and Utrecht. The WSIs in these two institutes demonstrate same type of tissue with different stain appearance due to variance in slide preparation. In this study, 100,000 256$\times$256 patches are randomly extracted from WSIs generated in Radbound institute for training, and another 20,000 256$\times$256 patches from Radbound are used for validation. The test set contains 80,000 256$\times$256 patches randomly extracted from Utrecht WSIs. Tumor patches are extracted from tumor regions in tumor slides and are treated as positive samples. Normal patches are extracted from non-tumor and non-background regions in normal slides and are treated as negative samples. The number of positive and negative patches are the same in all the training, validation and testing set.
	\subsubsection{Experimental Setup:} To start with, we use SGD to optimize the diagnosis net with a learning rate of $10^{-3}$, batch size of 8 on training set for 100 epochs. Then we use SGD with a learning rate of $10^{-4}$ and a batch size of 4 to train the GAN based style transfer model on the training set for 60 epochs. Hyper-parameters (i.e. weights of different loss functions) are turned using the validation set. And we finally choose $\alpha = 0.2 ,\beta=0.3 ,\gamma=0.5$.
	
	 To evaluate the proposed method, two experiments are performed. First, we execute style transfer on test images from Utrecht institute and examine generated images qualitatively in terms of stain resemblance to template images, color consistency, and preservation of histopathological information. Second, we evaluate the learning efficiency of our model and investigate the effectiveness of the SSIM and DSCSI based loss function in GAN learning. To this end, we record the image reconstruction loss in training to trace the optimization procedure. 
	
	For comparison, we also apply above experiments to prior arts by Cho\cite{c0}, Zanjani\cite{d0}, Janowczyk\cite{h0}, Li\cite{g0} and Macenko\cite{f0}. Among these methods, Cho’s method and the proposed methods have similar architectures. Based on GAN, Cho's method generates normalized images from the grayscale versions of query color images and uses MSE as reconstruction loss. Cho and Zanjani both took advantage of deep learning to generate color-normalized images. Note that compared to deep learning methods which learn stain and pattern features in the whole dataset, the last three methods use one slide as reference. For fair comparison, for the last three non-deep learning methods, we randomly choose patches from the training set as reference and execute color normalization over test set. We repeat the experiment ten times and report the average results.
	
	\subsubsection{Results and Discussions:}
	
	\begin{figure}[t]
		\centering
		\includegraphics[width=1\linewidth]{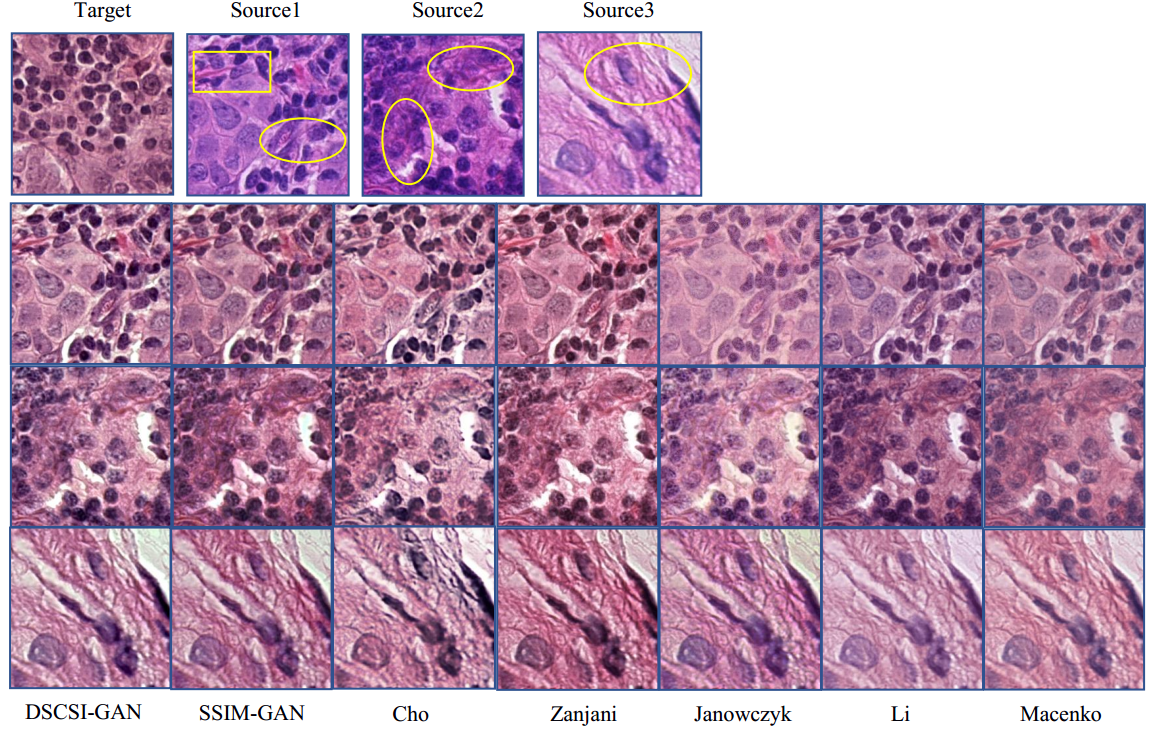}
		\caption{Stain normalization results of source images by different normalization methods. Target image in first row is sampled from training set as reference image and Source images 1-3 are sampled from testing set for visualization. Images from row2-4 respectively show the outputs of source images 1,2,3 generated by different methods. Images generated by last three non-deep learning methods appear quite different from target image. In the source images, circular and rectangular markers respectively show the areas that have problems of color inconsistency and histological information loss. The images generated by DSCSI-GAN and SSIM-GAN have comparatively more consistent stain appearance and the former ones have more clear histological structure.}
		\label{experiment}
	\end{figure}
	
\begin{figure}
	\centering
	\includegraphics[width=4.8in]{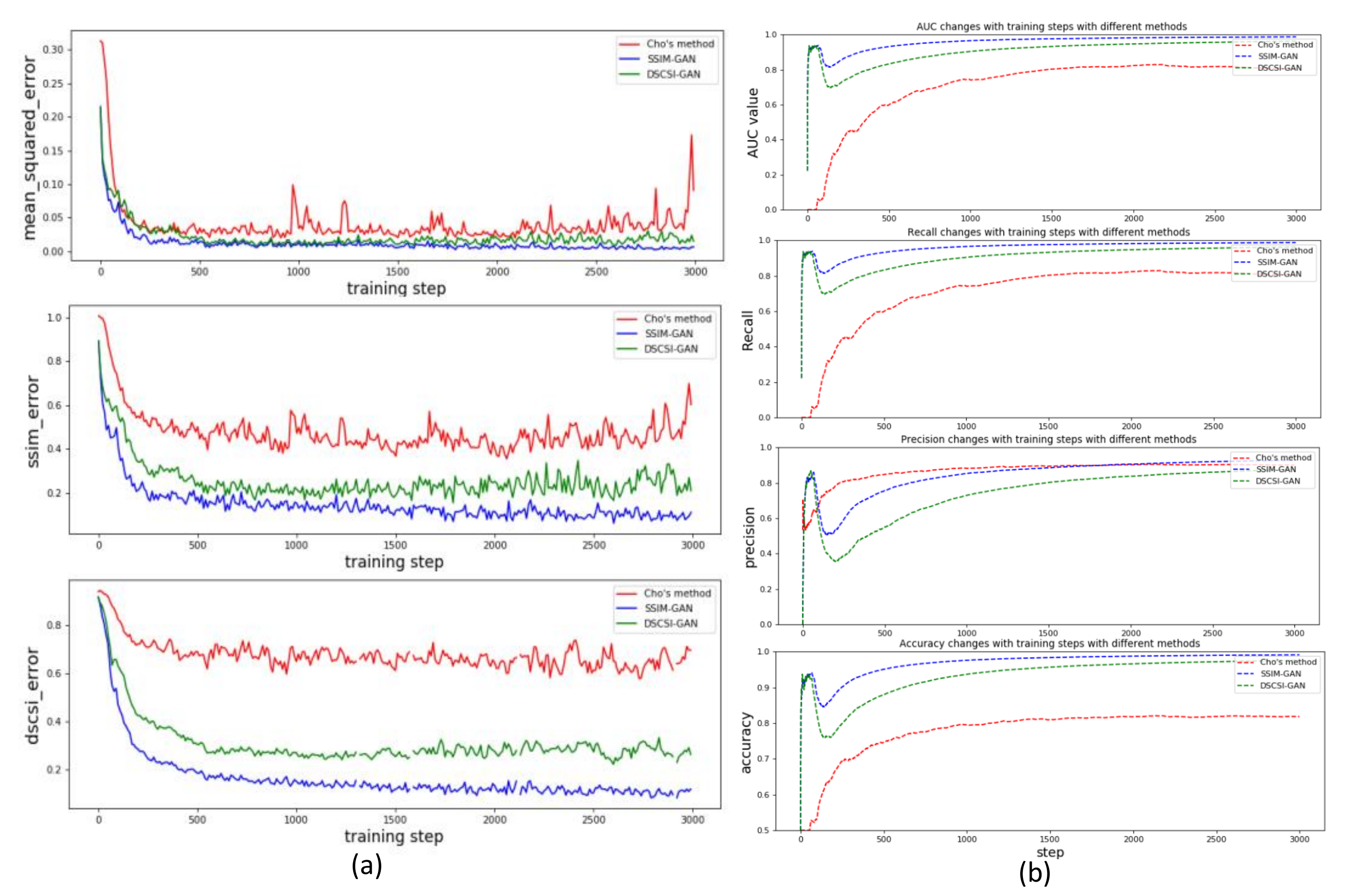}
	\caption{(a)Decrease of reconstruction errors in training. (b)Learning curve in the context of histopathology image diagnosis.}
	\label{curve}
\end{figure}
	
	First, we compare different normalization methods by visual examination. As you can see in Fig.~\ref{experiment}, the target image sampled from training set serves as a comparison reference. Source images are sampled from test set and the below are the stain normalized results of each method. Comparing all the generated images of three samples, we observe that the first four methods obtain better results than the last three non-deep learning methods. Among the four deep learning methods, images generated by the DSCSI-GAN and SSIM-GAN methods have comparatively more consistent stain appearance and the former has clearer histopathological structure. For instance, for Cho's result, the pink tissue marked by yellow rectangular in Source 1 disappears and tissues marked by yellow circular have stains inconsistent with the sources. The disadvantage in Cho's method is attributed to the discard of color information in stain transfer and the misuse of mean-squared-error(MSE) for image reconstruction. Images generated by Zanjani’s method seem better than those from Cho’s method and have consistent stain style as well as darker stain in cell areas.
	
	Second, we investigate the effectiveness of the proposed structure-preserving reconstruction loss on GAN's learning efficiency. 
	Under the same experimental setting, we record the learning efficiency by SSIM based error, DSCSI based error and MSE in Cho's method, shown in FIG.\ref{curve}(a). Compared with Cho's method where MSE value fluctuates severely, SSIM-GAN and DSCSI-GAN converge more quickly with steady decline. We also use diagnosis network to classify tumor and normal images and compute AUC, Precision, Recall and Accuracy score during training. The changes of four scores are shown in FIG. \ref{curve}(b). The AUC value, Recall and Accuracy increase more quickly and reach highest values in SSIM-GAN than others. Cho's model is the slowest among the three. 
	These results demonstrate higher efficiency and better optimization performance of the proposed structure preserving loss in this study.

	In summary, our methods have two advantages. First, opposed to traditional stain normalization methods, our model learn color distribution referencing to the whole training set not a single template. This helps to reduce the sensitivity of the normalization method to a particular training case. Second, we propose the use of structural similarity metrics, SSIM and DSCSI, as measurements of reconstruction error in GAN's training and induce the generator to generate high quality images. Since the introduced metrics drive the generator to learn structural content in the hue, chroma, lightness domains effectively, the learning efficiency is high in training.  
	
	\section{Conclusion}
	\label{conclusionsec}
	This work presented two stain style transfer models to solve the stain variation problem in computational histopathology. We took advantage of GAN that could learn stain distributions from a template dataset and obtain strong generalization capability to transfer the stain pattern to other datasets. In the proposed methods, we exploited SSIM and DSCSI to construct the reconstruction loss which prompt the model to maintain texture, structure, and color features in original images. 
	Extensive experimentation on publicly available dataset demonstrates that the proposed framework outperforms prior stain normalization solutions in generating stain-consistent images, preserving histopathological information, and obtaining high training efficiency.
	\clearpage
	
\bibliographystyle{IEEEbib}
\bibliography{ref_miccai}

\begin{thebibliography}{10}

\bibitem{b5}
Hans~O Lyon, Andre P.~De Leenheer, Richard~W. Horobin, Willy E~E Lambert,
  E.~K.~W. Schulte, B.~M.~Van Liedekerke, and D.~H. Wittekind,
\newblock ``Standardization of reagents and methods used in cytological and
  histological practice with emphasis on dyes, stains and chromogenic
  reagents,''
\newblock {\em The Histochemical Journal}, vol. 26, pp. 533--544, 1994.

\bibitem{i8}
Erik Reinhard, Michael Ashikhmin, Bruce Gooch, and Peter Shirley,
\newblock ``Color transfer between images,''
\newblock {\em IEEE Computer Graphics and Applications}, vol. 21, pp. 34--41,
  2001.

\bibitem{g8}
Ali Tabesh, Mikhail Teverovskiy, Ho-Yuen Pang, Vinay~P. Kumar, David Verbel,
  Angeliki Kotsianti, and Olivier Saidi,
\newblock ``Multifeature prostate cancer diagnosis and gleason grading of
  histological images,''
\newblock {\em IEEE Transactions on Medical Imaging}, vol. 26, pp. 1366--1378,
  2007.

\bibitem{h0}
Andrew Janowczyk, Ajay Basavanhally, and Anant Madabhushi,
\newblock ``Stain normalization using sparse autoencoders (stanosa):
  Application to digital pathology.,''
\newblock {\em Computerized medical imaging and graphics : the official journal
  of the Computerized Medical Imaging Society}, vol. 57, pp. 50--61, 2017 Apr
  2017.

\bibitem{b9}
Adnan~Mujahid Khan, Nasir~M. Rajpoot, Darren Treanor, and Derek~R. Magee,
\newblock ``A nonlinear mapping approach to stain normalization in digital
  histopathology images using image-specific color deconvolution,''
\newblock {\em IEEE Transactions on Biomedical Engineering}, vol. 61, pp.
  1729--1738, 2014.

\bibitem{b22}
Abhishek Vahadane, Tingying Peng, Amit Sethi, Shadi Albarqouni, Lichao Wang,
  Maximilian Baust, Katja Steiger, Anna~Melissa Schlitter, Irene Esposito, and
  Nassir Navab,
\newblock ``Structure-preserving color normalization and sparse stain
  separation for histological images,''
\newblock {\em IEEE Transactions on Medical Imaging}, vol. 35, pp. 1962--1971,
  2016.

\bibitem{b23}
Devrim Onder, Selen Zengin, and S{\"u}len Sarıoğlu,
\newblock ``A review on color normalization and color deconvolution methods in
  histopathology.,''
\newblock {\em Applied immunohistochemistry \& molecular morphology : AIMM},
  vol. 22 10, pp. 713--9, 2014.

\bibitem{k90}
Neslihan Bayramoglu, Mika Kaakinen, Lauri Eklund, and Janne Heikkil{\"a},
\newblock ``Towards virtual h\&e staining of hyperspectral lung histology
  images using conditional generative adversarial networks,''
\newblock {\em 2017 IEEE International Conference on Computer Vision Workshops
  (ICCVW)}, pp. 64--71, 2017.

\bibitem{c0}
Hyungjoo Cho, Sungbin Lim, Gunho Choi, and Hyunseok Min,
\newblock ``Neural stain-style transfer learning using gan for
  histopathological images,''
\newblock {\em CoRR}, vol. abs/1710.08543, 2017.

\bibitem{MICCAI2019}
Amal Lahiani, Nassir Navab~Shadi Albarqouni, and Eldad Klaiman,
\newblock ``Perceptual embedding consistency for seamless reconstruction of
  tilewise style transfer,''
\newblock in {\em MICCAI}, 2019.

\bibitem{ISBI2019}
M~Tarek Shaban, Christoph Baur, Nassir Navab, and Shadi Albarqouni,
\newblock ``Staingan: Stain style transfer for digital histological images,''
\newblock in {\em ISBI}, 2019.

\bibitem{MLMIR2019}
Shaojin Cai, Yuyang Xue, Qinquan Gao, Min Du, Gang Chen, Hejun Zhang, and Tong
  Tong,
\newblock ``Stain style transfer using transitive adversarial networks,''
\newblock in {\em MICCAI-MLMIR workshop}, 2019.

\bibitem{g21}
Arnout Ruifrok and Dennis Johnston,
\newblock ``Quantification of histochemical staining by color deconvolution,''
\newblock {\em Anal Quant Cytol Histol}, vol. 23, 01 2001.

\bibitem{p0}
Ian~J. Goodfellow, Jean Pouget-Abadie, Mehdi Mirza, Bing Xu, David
  Warde-Farley, Sherjil Ozair, Aaron~C. Courville, and Yoshua Bengio,
\newblock ``Generative adversarial nets,''
\newblock in {\em NIPS}, 2014.

\bibitem{e4}
Zhou Wang and A.~C. Bovik,
\newblock ``A universal image quality index,''
\newblock {\em IEEE Signal Processing Letters}, vol. 9, pp. 81--84, 2002.

\bibitem{lee2015towards}
Dohyoung Lee and Konstantinos~N Plataniotis,
\newblock ``Towards a full-reference quality assessment for color images using
  directional statistics,''
\newblock {\em IEEE Transactions on image processing}, vol. 24, no. 11, pp.
  3950--3965, 2015.

\bibitem{m0}
Olaf Ronneberger, Philipp Fischer, and Thomas Brox,
\newblock ``U-net: Convolutional networks for biomedical image segmentation,''
\newblock in {\em MICCAI}, 2015.

\bibitem{n0}
Kaiming He, Xiangyu Zhang, Shaoqing Ren, and Jian Sun,
\newblock ``Deep residual learning for image recognition,''
\newblock {\em 2016 IEEE Conference on Computer Vision and Pattern Recognition
  (CVPR)}, pp. 770--778, 2016.

\bibitem{Wang2004}
Z.~Wang, AC~Bovik, HR~Sheikh, and EP~Simoncelli,
\newblock ``Image quality assessment: From error visibility to structural
  similarity,''
\newblock {\em IEEE Transactions on Imaging Processing,}.

\bibitem{zhao2017loss}
Hang Zhao, Orazio Gallo, Iuri Frosio, and Jan Kautz,
\newblock ``Loss functions for image restoration with neural networks,''
\newblock {\em IEEE Transactions on Computational Imaging}, vol. 3, no. 1, pp.
  47--57, 2017.

\bibitem{o12}
Hang Zhao, Orazio Gallo, Iuri Frosio, and Jan Kautz,
\newblock ``Loss functions for image restoration with neural networks,''
\newblock {\em IEEE Transactions on Computational Imaging}, vol. 3, 2017.

\bibitem{zhang1997spatial}
Xuemei Zhang and Brian~A Wandell,
\newblock ``A spatial extension of cielab for digital color-image
  reproduction,''
\newblock {\em Journal of the society for information display}, vol. 5, no. 1,
  pp. 61--63, 1997.

\bibitem{dataset}
``Camelyon 2016 {https://camelyon16.grand-challenge.org},'' 2016.

\bibitem{d0}
Farhad~Ghazvinian Zanjani, Svitlana Zinger, Babak~Ehteshami Bejnordi, and
  Jeroen van~der Laak,
\newblock ``Histopathology stain-color normalization using deep generative
  models,''
\newblock 2018.

\bibitem{g0}
X.~Li and K.~N. Plataniotis,
\newblock ``A complete color normalization approach to histopathology images
  using color cues computed from saturation-weighted statistics,''
\newblock {\em IEEE Transactions on Biomedical Engineering}, vol. 62, no. 7,
  pp. 1862--1873, July 2015.

\bibitem{f0}
Marc Macenko, Marc Niethammer, J.~S. Marron, David Borland, John~T. Woosley,
  Xiaojun Guan, Charles Schmitt, and Nancy~E. Thomas,
\newblock ``A method for normalizing histology slides for quantitative
  analysis,''
\newblock {\em 2009 IEEE International Symposium on Biomedical Imaging: From
  Nano to Macro}, pp. 1107--1110, 2009.

\end{thebibliography}
\end{document}